\documentclass{article}
\usepackage{amssymb}
\usepackage{amsmath,bm}
\usepackage{array}

\title{Classical analog of quantum Schwarzschild black hole: local vs global, and the mystery of $\log  3$}
\author{Victor Berezin}
\date{
\small Institute for Nuclear Research, Russian Academy of Sciences, 
\\ 60th October Anniversary pr., 7-a, 117312 Moscow, Russia \\
{ \ } \\
and \\
{ \ } \\
Institut des Hautes \'Etudes Scientifiques, 35 route de Chartres, 
\\ F-91440 Bures-sur-Yvette, France 
{ \ } \\
{ \ } \\
e-mail: berezin@ms2.inr.ac.ru
}

\begin{document}

\maketitle

\vglue 2cm

\begin{abstract}
The model is built in which the main global properties of classical and quasi-classical black holes become local. These are the event horizon, ``no-hair'', temperature and entropy. Our construction is based on the features of a quantum collapse, discovered when studying some quantum black hole models. But our model is purely classical, and this allows to use selfconsistently the Einstein equations and classical (local) thermodynamics and explain in this way the ``$\log 3$''-puzzle.
\end{abstract}

\newpage

\section{Introduction and preliminaries}\label{sec1}

Classical definition of the black hole is based on the existence of the event horizon \cite{HE} -- the boundary of a space-time region from which the light cannot escape to infinity. The very notion of the event horizon is global and requires the knowledge of the whole history, both past and future.

\medbreak

Classical ``black hole has no hair'' \cite{RW} and is described by only few parameters, mass, Coulomb-like charge and angular momentum. The Schwarzschild black hole has only the mass, the Reissner-Nordstrom one -- mass and charge, the Kerr black one -- mass and angular momentum. The most general type -- Kerr-Newman black hole -- has all three parameters. This resembles the body in thermal equilibrium. The process of becoming bold is also global, its  duration, formally, is infinite, like the process of coming to thermal equilibrium. It goes through radiating of all possible (scalar, vector, spinor, tensor) perturbations and governed by Schroedinger-like wave equation, first derived in \cite{ReW}. The results of many numerical studies for a long period (two decades) were summarized in the book \cite{Ch}. It appeared that such perturbation modes have discrete spectra with complex frequencies $w$. They received the name ``quasi-normal modes'' and ``quasi-normal frequencies'', respectively \cite{Pr}. The imaginary parts are equidistant indicating that the decaying modes are radiating away in a manner reminiscent of the last pure dying tones (infinitely many overtones) of a ringing bell, and the higher the overtone, the shorter its lifetime. The real part of quasi-normal frequencies tends to some constant value which depends on the black hole type. For Schwarzschild black holes, we are interested in here,
\begin{equation}
\label{QNMN}
Gm \, w_n = 0.0437123 -  \frac{i}{4} \left( n+\frac{1}{2} \right) + O [(n+1)^{-1/2}] \, , \quad n \to \infty
\end{equation}
where $m$ is the mass, and $G$ is the Newton's constant. All that shows that black holes have some inherent frequency. Therefore, they are not ``dead'' but have some ``private life'', encoded in some features of their horizons. Evidently, this is also the global property because it does not depend on what is going on inside.

\medbreak

Investigation  of the processes near an event horizon showed that they can be reversible and irreversible like in thermodynamics \cite{C,CR}. The assimilation of a point (classical) particle by a (non extremal\footnote{If a black hole has more than one parameter, then for fixed value of other, than mass, parameters there exists the minimal value of mass (critical, or extreme), below which the event horizon (and, thus, the black hole itself) does not exist.}) black hole reversible if it is injected at the event horizon from a radial turning point of its motion. In this case the black hole (horizon) area remains unchanged, and the change in other parameters (mass, charge and angular momentum) can be undone by another suitable (reversible) process. In all other cases the horizon area $A$ increases. Thus, for classical black holes
\begin{equation}
\label{HA}
dA \geq 0 \, .
\end{equation}

\medbreak

The new area in black hole physics started with the seminal paper by J.D.~Bekenstein \cite{B}, where he presented serious physical arguments that the Schwarzschild black hole should be described by a certain amount of entropy which is proportional to the area of event horizon. Such a strict proportionality could appear to be playing games with symbols with only one parameter, black hole mass, but it was then confirmed by J.M.~Bardeen, B.~Carter and S.W.~Hawking \cite{BCH} who proved the four laws of thermodynamics for the general class of Kerr-Newman black holes. Moreover, it was shown that the role of the temperature is played by the surface gravity $\varkappa$ at the event horizon (up to some numerical factor), which is constant there. And only after discovering by S.W.~Hawking the black hole evaporation \cite{H} this thermodynamical analogy became the real physical phenomenon. He considered the quantum theory of massless scalar field on the Schwarzschild static space-time background and found that the specific boundary conditions -- only infalling waves in the vicinity of the horizon -- result in a thermal behavior of the wave functions and nonvanishing energy flow to the infinity. It appeared that the spectrum of such a radiation is Planckian with the temperature
\begin{equation}
\label{TH}
T_H = \frac{\varkappa_H}{2 \, \pi} \, ,
\end{equation}
where $\varkappa$ is the surface gravity at the event horizon. It follows then, that the black hole entropy is exactly one fourth of dimensionless horizon area
\begin{equation}
\label{E}
S = \frac{1}{4} \, \frac{A}{\ell^2_{p_{\ell}}} \, ,
\end{equation}
where $\ell_{p_{\ell}} = \sqrt{\frac{\hbar G}{c^3}} \sim 10^{-33} cm$ is the Planckian length ($\hbar$ is the Planck constant, $c$ is the speed of light, and $G$ is the Newton's gravitational constant). We will use the units $\hbar = c = k = 1$ ($k$ is the Boltzmann constant), so $\ell_{p_{\ell}} = \sqrt G$ and the Planckian mass is $m_{p_{\ell}} = \sqrt{\frac{\hbar c}{G}} = 1/\sqrt G \sim 10^{-5} gr$.

\medbreak

The nature of Hawking radiation and its black body spectrum lies in the nontrivial causal structure of the space-times containing black holes. The crucial point is the existence of the event horizons. The same takes place in the Rindler space-time. This space-time is obtained by transforming the two-dimensional Minkowski flat space-time from the ``ordinary'' coordinates $(t,x)$ and metric $ds^2 = dt^2 - dx^2$ related to the set of inertial observers, to the so-called Rindler coordinates $(\eta , \xi)$ and metric
$$
t = \frac{1}{a} \, e^{a\xi} \, \sinh a \eta \, , \quad x = \pm \frac{1}{a} \, e^{a\xi} \cosh a\eta \quad (x \gtrless 0) \, ,
$$
$$
-\infty < t < \infty \, , \quad -\infty < \xi < \infty
$$

\begin{equation}
\label{R}
ds^2 = e^{2a\xi} (d\eta^2 - d\eta^2) \, .
\end{equation}
Thus, the Rindler space-time is static and locally flat but differs from the two-dimensional Minkowski space-time globally, because it covers only one half of the latter and, in addition, possesses the event horizon at $t = \pm \, x$ ($\eta = \pm \, \infty$, $\xi = {\rm const}$). The Rindler observers at $\xi = {\rm const}$ are uniformly accelerated. The norm of the acceleration vector $a^{\mu}$ equals
\begin{equation}
\label{acc}
\alpha = \sqrt{-a^{\mu} \, a_{\mu}} = ae^{-a\xi} \, .
\end{equation}
Considering a quantum scalar field in the Rindler space-time, W.G.~Unruh found \cite{U}, in fact, the finite temperature quantum field theory with the temperature
\begin{equation}
\label{UT}
T_U = \frac{a}{2 \, \pi} \, .
\end{equation}
We see that this temperature is proportional to the acceleration of the Rindler observer sitting a $\xi = 0$ with $g_{00} = 1$. But, all of them are equivalent (we can always shift the spatial coordinate $\xi \to \xi - \xi_0$). The temperature is not an invariant, but it is a temporal component of a heat vector. This means that each observer measures the Unruh temperature when using its proper time $\tau$ ($ds = d\tau$). If the same observer uses the local clocks that show the local time $t$ ($ds = \sqrt{g_{00}} \, dt$), the local temperature measured by him equals
\begin{equation}
\label{LUT}
T_{\rm loc} = \frac{T_U}{\sqrt{g_{00}}} = \frac{a}{2 \, \pi} \, e^{-a\xi} = \frac{\alpha}{2 \, \pi} \, ,
\end{equation}
which is proportional to the local acceleration $\alpha$. The very fact that the uniformly accelerated observer ($=$ detector) will detect the real particles in the vacuum, was known to people doing quantum electrodynamics long ago. It was understood as a change of a vacuum state due to the external forces that cause such an acceleration. The same happens in the space-time with event horizons. But that the spectrum is thermal appeared to be new and purely relativistic feature. We know from the university course of thermodynamics (see, e.g. \cite{LL}) that the condition for thermal equilibrium in static space-times is $T_{\rm loc} \, \sqrt{g_{00}} = {\rm const}$. Thus, all the Rindler observers are in thermal equilibrium with each other. Is the Rindler space-time unique in this sense? To answer, let us consider some general two-dimensional static space-time with a metric
\begin{equation}
\label{tdm}
ds^2 = e^{\nu} \, dt^2 - d \rho^2 = e^{\nu} \, dt^2 - e^{\lambda} \, dq^2 \, .
\end{equation}
In the Rindler case $\rho = \frac{1}{a} \, e^{a \xi}$, $e^{\nu} = a^2  \rho^2 = g_{00}$. The static observer undergoes a constant acceleration with the invariant $\alpha = \frac{1}{2} \left\vert \frac{d\nu}{d\rho} \right\vert = \frac{1}{2} \left\vert \frac{d\nu}{dq} \right\vert e^{-\frac{\lambda}{2}}$, and the (now local) Rindler parameter $a(\rho)$, which is called ``the surface gravity $\varkappa$'', equals
\begin{equation}
\label{sgr}
\varkappa = \frac{1}{2} \left\vert \frac{d\nu}{dq} \right\vert e^{\frac{\nu - \lambda}{2}} = \frac{1}{2} \left\vert \frac{d\nu}{d\rho} \right\vert e^{\frac{\nu}{2}} \, .
\end{equation}
The thermal equilibrium requires $\varkappa = {\rm const}$, therefore, $g_{00} = C\rho^2$, and this proves that the Rindler space-time is the only one which static observers are in the mutual thermal equilibrium.

\medbreak

By the Einstein equivalence principle we can extend all we learned studying Rindler space-times, to the static gravitational fields, especially to the spherically symmetric ones, because after fixing spherical angles $\theta$ and $\varphi$ the latter become, in fact, the two-dimensional pseudo-surfaces. Of course, in general these surfaces are curved, the equivalence principle holds only locally, and the static observers will not be in thermal equilibrium with each other. Such a temperature is observer-dependent and cannot be considered as an intrinsic property of a given space-time. But for the black hole space-times the position of the event horizon is absolute and does not depend on the observer. So, its temperature does serve an important characteristic of space-time itself. To know the temperature we just need to compute the surface gravity value at the event horizon, $\varkappa_H$. For the Schwarzschild black hole with the famous metric
\begin{eqnarray}
\label{Sm}
ds^2 &= &F \, dt^2 - \frac{1}{F} \, dr^2 - r^2 (d\theta^2 + \sin^2 \theta \, d\varphi^2) \, , \nonumber \\
F &= &1-\frac{2 \, Gm}{r} \, ,
\end{eqnarray}
where $m$ is the black hole mass, and $r$ is the radius of a sphere (in that sense that its area equals $4 \, \pi \, r^2$), the horizon is located at the Schwarzschild radius $r_g = 2 \, Gm$, and the surface gravity equals
\begin{equation}
\label{sgrs}
\varkappa_H = \frac{1}{2} \left\vert \frac{d\nu}{dr} \right\vert \, e^{\frac{\nu-\lambda}{2}} = \frac{1}{2} \, F' (r_H) = \frac{Gm}{r^2} \biggl\vert_{r_g} = \frac{1}{4 \, Gm} \, .
\end{equation}
Therefore, the Hawking temperature is just the Unruh temperature at the event horizon measured by distant observers (at infinity). The same is true also for Kerr-Newman black holes. Note that outside the event horizon $r > r_g$ the Schwarzschild observers are not in thermal equilibrium with each other, and this is a thermodynamical explanation of the Hawking radiation and, thus, evaporation of black holes. It should be stressed that both the black hole temperature and entropy are global features because their very appearance is due to the existence of the event horizon.

\medbreak

Evaporating, black holes become smaller and smaller and will reach eventually a Planckian size where the still unknown quantum gravity should play an important role. Since the radiation is quantized, the black hole mass have to be quantized as well. Of course the relation is not direct because a black hole is not necessarily transformed into black hole again, but the new black hole will eventually be formed only by radiation. To the black hole mass there may contribute not only the rest masses and kinetic energy of particles, including the total angular momentum, but also Coulomb and magnetic energies of their electric and gauge charges and all kinds of other physical fields confined under the event horizon. But the common feature for all types of black holes is their entropy with its universal relation (\ref{E}) to the horizon area. Thus, the black hole quantization means the quantization of its entropy. Moreover, the thermodynamical description is possible only if the jump in the temperature due to quantization of mass, charge and angular momentum during black hole evaporation is negligible compared to its absolute value, while the notion of the entropy as a measure of the information, hidden or ignored, is still valid. This latter feature gives rise to common believe that the black hole quasiclassical quantization can shade light on the structure of the future full quantum gravity, or, at least, will provide us with some selection rules in the  attempts to construct such a theory. The quantization of a black hole as whole was proposed long ago by  J.~Bekenstein \cite{B1}. The idea was based on the remarkable observation that the horizon area of non-extremal black holes behaves as a classical adiabatic invariant. The Bohr-Sommerfeld quantization rule then predicts the equidistant spectrum for the horizon area and thus, for the black hole entropy. The gedanken experiments show that, due to the quantum effects, the minimal increase in the horizon area in the processes of capturing a neutral \cite{B2} or electrically charged \cite{DR} particle approximately equals
\begin{equation}
\label{Amin}
\Delta \, A_{\rm min} \approx 4 \, \ell^2_{p_{\ell}} \, .
\end{equation}
This suggests for the black hole entropy
\begin{equation}
\label{edsp}
S_{BH} = \gamma_0 \, N \, , \quad N = 1,2,\ldots
\end{equation}
where $\gamma_0$ is of order of unity. In their famous work on the black hole spectroscopy J.D.~Bekenstein and V.F.~Mukhanov \cite{BM} related the black hole entropy to the number $g_n$ of microstates that corresponds to the particular external macrostate through the well-known formula in statistical physics
$g_n = \exp [S_{BH} (n)]$, i.e., $g_n$ is the degeneracy of the $n$-th area eigenvalue. Since $g_n$ should be integer, they deduce that
\begin{equation}
\label{gz}
\gamma_0 = \log k \, , \quad k=2,3,\ldots
\end{equation}
In the spirit of the information theory and the famous claim by J.A.~Wheeler ``It from Bit'' the value of $\log 2$ seems most suitable one.

\medbreak

The logarithmic behavior of the spacing coefficient $\gamma_0$ comes also from the Loop Quantum Gravity. It was shown in \cite{ABCK}, \cite{ABK}, that the entropy of the Schwarzschild black hole is proportional to the horizon area as well as a numerical constant called the Barbero-Immirzi parameter. To fit the Bekenstein-Hawking relation (\ref{E}) and the possible value for $\gamma_0$ (\ref{gz}) this parameter should equal $\log 2 / (\pi \sqrt 3)$ if the fundamental group in LQG is $SU(2)$, and $\log 3 / (2\pi \sqrt 2)$ if it is $SU(3)$. The choice of the value for $\gamma_0$ leads to minimal possible change in the black hole mass. S.~Hod \cite{Hod}, using famous Bohr's correspondence principle (1923): ``Transition frequencies at large quantum numbers should equal classical oscillation frequencies'', deduced that the real part of the complex quasi-normal frequencies for the Schwarzschild black hole should be proportional to $\log 3$. And, indeed,
\begin{equation}
\label{Rep}
Gm \, Re \, w = 0.0437123 = \frac{\log 3}{8 \, \pi} \, .
\end{equation}
The value of $\gamma_0$ as well as that of Barbero-Immirzi parameter and, thus, the choice of the fundamental group in LQG, must be universal. Therefore, it is not surprising that people tried to find some analytical methods for evaluating the quasi-normal frequencies for different types of black holes. By using rather sophisticated tools from the general theory of ordinary differential equations, L.~Molt and A.~Neitzke showed \cite{M}, \cite{MN} that for the scalar and tensor perturbations around Schwarzschild black holes the value $\log 3$ is exact. For more general types of black holes the corresponding calculations were fulfilled in \cite{RS}. It appeared that the simple value $\log 3$ for the spacing coefficient $\gamma_0$ is by no means universal, but exceptional. That is why we use the expression ``the mystery of $\log 3$''.

\medbreak

Below we construct a model which is not really a black hole, but possesses its main features. It has an event horizon -- but local, the temperature -- but local. Then, we develop the local thermodynamics for such a model and show how the mystery of $\log 3$ can be solved.

\section{Quantum thin shells}
\setcounter{equation}{0}

The geodesically complete Schwarzschild space-time has a geometry of non-transversable wormhole (it is also called an eternal black hole). There are two asymptotically flat regions with spatial infinities connected by the Einstein-Rosen bridge (the throat). Two sides of the bridge are causally disconnected and separated by (past and future) event horizons. The gravitating source is concentrated at two space-like (momentarily existing) singular surfaces at zero radius. In a sense, there is nothing to quantize. To get physically meaningful results we need to introduce some dynamical gravitating source. The simplest generalization of the point mass is the spherically symmetric self-gravitating thin dust shell.

\medbreak

The theory of thin shells in General Relativity was developed by W.~Israel \cite{I} and applied then to various problems by many authors. In this approach the whole space-time is divided into three parts: interior (in), exterior (out) and the hypersurface in-between, called ``shell'' or ``brane'' (short for ``membrane''). The matter source on the shell is proportional to $\delta$-function and described by a surface energy momentum tensor. The equations of motion are obtained by integrating the Einstein equations along the normal direction to the shell from ``in'' to ``out''. In the case of spherical symmetry the only dynamical variable is the shell radius $\rho (\tau)$ as a function of the proper time of the observer sitting on the shell. Our shell consists of dust. This means that it is characterized by the bare mass $M$ (which is just the sum of masses of the constituent particles), and the only non-zero component of the surface energy momentum tensor is $S_0^0 = \frac{M}{4 \, \pi \, \rho^2}$. Thus, we need only one equation, and this is the energy constraint $\left( {0 \atop 0} \right)$-Einstein equation. We assume that both inside and outside the shell there are vacuum Schwarzschild metrics with the masses, correspondingly $m_{\rm in}$ and $m_{\rm out}$. Then the shell equation reads as follows (a dot denotes the time derivative)
\begin{equation}
\label{sheq}
\sigma_{\rm in} \sqrt{\dot\rho^2 + F_{\rm in}} - \sigma_{\rm out} \sqrt{\dot\rho^2 + F_{\rm out}} = \frac{GM}{\rho}
\end{equation}
$$
F = 1 - \frac{2 \, Gm}{\rho} \, .
$$
Here $\sigma = \pm \, 1$ is the sign function that indicates whether the radii increase in the normal outward direction $(\sigma = +1)$, or they decrease $(\sigma = -1)$. When $\sigma_{\rm in} = -1$, this means that in the interior we have a semiclosed world, then for $\sigma_{\rm out} = -1$ we obtain $M < 0$, i.e., the shell consists of particles with negative mass what is too exotic. If $\sigma_{\rm out} = -1$ and $M > 0$, then in this case outside the shell we would have no spatial infinity at all but the curvature singularity at zero radius instead. We exclude the case $\sigma_{\rm in} = -1$ for physical reasons. Therefore, the global geometry of the exterior region is determined by the sign of $\sigma_{\rm out} = \sigma$. The sign of $\sigma$ is dynamically changed at the point where $\dot\sigma^2 + F_{\rm out} = 0$, so $F_{\rm out} < 0$, that is, beyond the event horizon $r_g = 2 \, Gm_{\rm out}$, but outside it is a constant of motion. Thus, the value of $\sigma$ distinguishes between two different types of shells. If $\sigma = +1$ outside the horizon, the shell moves on ``our'' side of the Einstein-Rosen bridge. In the opposite case it forms the semiclosed world and does not appear at all in ``our'' part of the whole space-time manifold. This depends on the relation between the mass inside $m_{\rm in}$, the bare mass $M$ of the shell itself and on the total mass of the system $m$ that includes the gravitational mass defect and is determined by the shell initial data (position and velocity). It is not difficult to show that
\begin{eqnarray}
\label{SS}
\sigma = +1 &{\rm if} &\frac{\Delta m}{M} > \frac{1}{2} \left( \sqrt{\frac{m_{\rm in}^2}{M^2} + 1} - \frac{m_{\rm in}}{M} \right) \nonumber \\
\sigma = -1 &{\rm if} &\frac{\Delta m}{M} < \frac{1}{2} \left( \sqrt{\frac{m_{\rm in}^2}{M^2} + 1} - \frac{m_{\rm in}}{M} \right) \, .
\end{eqnarray}
Here $\Delta m$ is the total mass of the shell, including the mass defect, $\Delta m = m_{\rm out} - m_{\rm in}$. In what follows we will be interested in bound motion only, so $\frac{\Delta m}{M} < 1$. The above inequalities were derived using the explicit expression for the turning point $\rho_0$. In the case of quantum shells there is no trajectories and no turning point, the latter being replaced by a principal quantum number $n$. So, the type of the shell should be determined in another way. At the turning point $\dot\rho = 0$ and our shell equation becomes
\begin{equation}
\label{etp}
\sqrt{1-\frac{2 \, Gm_{\rm in}}{\rho_0}} - \sigma \sqrt{1 - \frac{2 \, Gm_{\rm out}}{\rho_0}} = \frac{GM}{\rho_0} \, .
\end{equation}
We see that the sign of a derivative $\partial \Delta m / \partial M$ for fixed $m_{\rm in}$ and $\rho_0$ coincides with the sign of $\sigma$. For quantum shell we, thus, have
\begin{equation}
\label{qs}
\sigma = {\rm sign} \, \frac{\partial \Delta m}{\partial M} \biggl\vert_{n,m_{\rm in}} \, .
\end{equation}

The first attempt to quantize a spherically symmetric thin dust shell, using the squared version of Eqn.~(\ref{sheq}) and for flat $(m_{\rm in} = 0)$ interior, were made in \cite{BKKT}. The total mass $m_{\rm out} = m_{\rm tot} = \Delta m$ was considered as the energy $E$ of the system, and the squared equation -- as the pre-Hamiltonian $E(\dot\rho , \rho)$, from which the conjugate momentum $p$ and the Hamiltonian $H(p,\rho)$ was derived. In this approach there remained no trace of $\sigma$ and, thus, no trace of the non-trivial causal structure of the geodesically complete Schwarzschild space-time. The same straightforward quantization of the original equation, again for $m_{\rm in} = 0$, was fulfilled in \cite{Ber}. The result was the following Schroedinger-like equation in finite differences
\begin{equation}
\label{me}
\Psi (S + i\zeta) + \Psi (S-i\zeta) = \frac{2-\frac{1}{\sqrt S} - \frac{M^2}{4 \, m^2 S}}{\sigma \sqrt{1-\frac{1}{\sqrt S}}} \, \Psi (S)
\end{equation}
where $m=m_{\rm out} = m_{\rm tot}$, $M$ is the bare mass of the shell, $S = \frac{\rho}{4 \, G^2 m^2}$, $\zeta = \frac{1}{2 \, G m^2} = \frac{m^2_{p_{\ell}}}{2 \, m^2}$. This equation in the above-written form is valid only outside the horizon. But, if we consider $\sqrt{ \ }$ as a complex function $( \ )^{1/2}$ of one complex variable, then it can be continued beyond the horizon into two other regions (of inevitable contraction -- black hole region, and of inevitable expansion -- white hole region) of the whole Schwarzschild space-time, the horizon being at the branching point $S=1$. The physically acceptable solutions for the wave function $\Psi$ should be exponentially damped in both asymptotically flat regions (at both spatial infinities). Then, for large enough black holes $(\zeta \ll 1)$ it appears that for the appropriate choice of the pass around the branching point, the leading term in the region of inevitable contraction is an ingoing wave, while in the region of inevitable expansion it is an outgoing wave, as it should be expected for the quasi-classical motion of the shell.

\medskip

The ADM-formalism for spherically symmetric space-times with a thin shell was developed in \cite{BBN} whith the Wheeler-DeWitt (Schroedinger-like) equation of the same type as Eqn.~(\ref{me}), but for general case $m_{\rm in} \ne 0$:
\begin{eqnarray}
\label{oe}
\Psi (S + i\zeta) + \Psi (S - i\zeta) &= &\frac{F_{\rm in} + F_{\rm out} - \frac{M^2}{4 \, m^2 S}}{\sigma \sqrt{F_{\rm in}} \, \sqrt{F_{\rm out}}} \, \Psi (S) \, , \nonumber \\
F \left( {{\rm in} \atop {\rm out}} \right) &= &1 - \frac{2 \, G m \left( {{\rm in} \atop {\rm out}} \right)}{\rho} \, .
\end{eqnarray}
Note that the shift in the argument is purely imaginary. Therefore, the ``good'' solutions should be analytical functions. Besides, there are branching points. So, the wave functions should be analytical on some Riemann surface. The physical reason for this is the following. In quantum theory there are no trajectories. Thus, if a shell has parameters $\Delta m$ and $M$ corresponding, say, to $\sigma = +1$ and to motion on ``our'' side of the Einstein-Rosen bridge, its wave function is nonzero on the other side as well and ``feels'' both infinities where we have to impose the appropriate boundary conditions. Comparing the behavior of the solutions to the Eqn.~(\ref{oe}) in vicinities of singular points (infinities and singularities) and around branching points, the following quantum conditions were found for a discrete mass spectrum in the case of bound motion:
\begin{eqnarray}
\label{dsp}
\frac{2 \, (\Delta m)^2 - M^2}{\sqrt{M^2 - (\Delta m)^2}} &= &\frac{2 \, m^2_{p_{\ell}}}{\Delta m + m_{\rm in}} \, n \nonumber \\
M^2 - (\Delta m)^2 &= &2 \, (1+2p) \, m^2_{p_{\ell}} \, ,
\end{eqnarray}
where $n$ and $p$ are integers. The appearance of two quantum numbers instead of one as in conventional quantum mechanics is due to the nontrivial causal structure of the complete Schwarzschild manifold. The principal quantum number $n$ comes from the boundary condition at ``our'' infinity, while the new, second, quantum number $p$ -- from the infinity on the other side of the Einstein-Rosen bridge.

\medbreak 

Wave functions corresponding to this discrete spectrum form the two-para\-meter family $\Psi_{n,p}$. But for the quantum Schwarzschild black holes we expect a one-parameter family of solutions because quantum black holes should have ``no hairs'', otherwise there will be no smooth classical limit. This means that our spectrum is not a quantum black hole spectrum, and corresponding quantum shells do not collapse, like an electron in hydrogen atom. Physically, it is quite understandable, because the radiation was not included into consideration. Since two quantum numbers determine completely the parameters of the shell, $\Delta m$ and $M$, for fixed $m_{\rm in}$, the quantum collapse goes through decrease in $m_{\rm out}$, due to radiation, and increase in $m_{\rm in}$, due to creation new shells, thus diminishing $\Delta m$. Of course, our picture is extremely simplified, but in the real collapse, during which there can be deviations from the strict spherical symmetry, the tendency is the same. This process can go in many different ways, so, the quantum collapse is accompanied with the loss of information, thus converting an initially pure quantum state into some thermal mixed one. But how could quantum collapse be stopped? The natural limit is the crossing of the Einstein-Rosen bridge, since such a transition requires (at least in a quasi-classical regime) insertion of infinitely large volume, with, of course, zero probability. Computer simulations show that the process of quantum collapse for our shells stops when the principal quantum number becomes zero, $n=0$.

\medbreak

The point $n=0$ in our spectrum is very special. In this case the shell does not ``feel'' not only the outer region (what is natural for the spherical configuration) but it does not know anything about what is going on inside. It ``feels'' only itself. Such a situation reminds the ``no hair'' property of a classical black hole. Finally, when all the shells (both the primary one and newly born) are in the corresponding states $n_i = 0$, the whole system does not ``remember'' its own history. Then it is this ``no-memory'' state that can be called ``the quantum black hole''. Note that the total masses of all the shells obey the relation
\begin{equation}
\label{sqrt2}
\Delta \, m_i = \frac{1}{\sqrt 2} \, M_i \, .
\end{equation}
The subsequent quantum Hawking radiation can proceed via some collective excitations.

\newpage

\section{Classical analog of quantum Schwarzschild \\ black hole$^2$}
\setcounter{equation}{0}
\footnotetext[2]{The main part (up to the last paragraph) of this section was elaborated in \cite{Ber1}.}

The final state of quantum gravitational collapse can be viewed as some stationary matter distribution. Therefore, we may hope that for massive enough quantum black hole such a distribution is described approximately by a classical static spherically symmetric perfect fluid with energy density $\varepsilon$ and (effective) pressure $p$ obeying classical Einstein equations. This is what we call a classical analog of a quantum black hole. Of course, such a distribution has to be very specific. To study its main features, let us consider the situation in more details.

\medbreak

Any static spherically symmetric metric can be written in the form
\begin{equation}
\label{ssm}
ds^2 = e^{\nu} \, dt^2 - e^{\lambda} \, dr^2 - r^2 (d\theta^2 + \sin^2  \theta \, d\varphi^2) \, .
\end{equation}
Here $r$ is the radius of a sphere with the area $A = 4 \, \pi \, r^2$, $\nu = \nu (r)$, $\lambda = \lambda (r)$. The Einstein equations are (prime denotes differentiation in $r$):
\begin{eqnarray}
\label{ee1}
-e^{-\lambda} \left( \frac{1}{r^2} - \frac{\lambda'}{r} \right) + \frac{1}{r^2} &= &8 \, \pi \, G \varepsilon \, , \nonumber \\
- e^{-\lambda} \left( \frac{1}{r^2} + \frac{\nu'}{r} \right) + \frac{1}{r^2} &= &-8 \, \pi \, G p \, , \nonumber \\
-\frac{1}{2} \left( \nu'' + \frac{\nu'^2}{2} + \frac{\nu' - \lambda'}{r} - \frac{\nu' \lambda'}{2} \right) &= &-8 \, \pi \, G p \, .
\end{eqnarray}
We see that there are three equations for four unknown functions. But, even we would know an equation of state for our perfect fluid, $p=p(\varepsilon)$, the closed (formally) set of equations would have too many solutions. We need, therefore, some selection rules in order to single out the classical analog of quantum black hole. Surely, the ``no hair'' feature should be the main criterium. Thus, we have to adjust our previous definition of the ``no-memory'' state to the case of a continuum matter distribution.

\medbreak

For this, let us integrate the first of Eqns.~(\ref{ee1}):
\begin{equation}
\label{int}
e^{-\lambda} = 1 - \frac{2 \, G m(r)}{r} \, ,
\end{equation}
where
\begin{equation}
\label{mf}
m(r) = 4 \, \pi \int_0^r \varepsilon \, \tilde r^2 \, d\tilde r
\end{equation}
is the mass function that should be identified with $m_{\rm in}$. Now, the ``no-memory'' principle is readily formulated as the requirement,  that $m(r) = ar$, i.e.$^3$\footnotetext[3]{The same type of energy density distribution was considered almost 50 years ago by Y.B.~Zel'dovich, but for quite different purposes \cite{Z}.},
\begin{eqnarray}
\label{nm}
e^{-\lambda} &= &1-2 \, Ga = {\rm const}, \nonumber \\
\varepsilon &= &\frac{a}{4 \, \pi \, r^2} \, .
\end{eqnarray}
We can also introduce a bare mass function $M(r)$ (the mass of the system inside a sphere of radius $r$ without gravitational mass defect):
\begin{equation}
\label{bmf}
M(r) = \int \varepsilon \, d \, V = 4 \, \pi \int_0^r \varepsilon \, e^{\frac{\lambda}{2}} \, \tilde r^2 \, d\tilde r = \frac{ar}{\sqrt{1-2 \, Ga}} \, .
\end{equation}
The remaining two equations can now be solved for $p(r)$ and $e^{\nu} (r)$. The general solution is rather complex, but the correct non-relativistic limit for the pressure $p(r)$ (we are to reproduce the famous equation for hydrostatic equilibrium) is given by only the following one-parameter family:
\begin{equation}
\label{p}
p(r) = \frac{b}{4 \, \pi \, r^2} \, ,
\end{equation}
where
\begin{equation}
\label{b}
b = \frac{1}{G} \left( 1-3 \, Ga - \sqrt{1-2 \, Ga} \, \sqrt{1-4 \, Ga} \right) \, .
\end{equation}
We see that the solution exists only for $a \leq \frac{1}{4 \, G}$, then $b \leq a$. The physical meaning of these inequalities is that the speed of sound cannot exceed the speed of light, $v_{\rm sound}^2 = \frac{b}{a} \leq 1 = c^2$, the equality being reached just for $a=b=\frac{1}{4 \, G}$. Finally, for the temporal metric coefficient $g_{00} = e^{\nu}$ we get
$$
e^{\nu} = C_0^2 \, r^{\frac{4b}{a+b}} = C_0^2 \, r^{2G \frac{a+b}{1-2Ga}} \, .
$$
Thus, demanding the ``no-memory'' feature and the existence of the correct non-relativistic limit, we obtained the two-parameter family of static solutions. But, we need a one parameter family, so we have to continue our search.

\medbreak

Evidently, the point $r=0$ is singular both for matter distribution and $g_{00}$ metric coefficient. To examine what kind of singularity we are dealing with, one should calculate the Riemann curvature tensor. It appears that for $b < a$ this tensor is, indeed, divergent at $r=0$. But, if $a=b=\frac{1}{4 \, G}$, we are witnessing a miracle, the (before) divergent components become zero. Thus, demanding, in addition to the previous two very natural requirements, the third one (also natural), namely, the absence of the real (curvature) singularity at $r=0$, we arrive at the following one-parameter family of solutions to the Einstein equations (\ref{ee1})
\begin{eqnarray}
\label{fs}
g_{00} &= &e^{\nu} = C_0^2 \, r^2 \, , \nonumber \\
g_{11} &= &-e^{\lambda} = -2 \, , \nonumber \\
\varepsilon &= &p = \frac{1}{16 \, \pi \, G r^2} \, .
\end{eqnarray}
So, the equation of state of our perfect fluid is the stiffest possible one. The constant of integration $C_0$ can be determined by matching the interior and exterior metrics at some boundary value of radius, $r=r_0$. Let us suppose that for $r > r_0$ the space-time is empty, so, the interior should be matched to the Schwarzschild metric with the mass parameter $m$. Of course, to compensate the jump in the pressure $\Delta p$ $( = p(r_0) = p_0)$ we must include in our model a surface tension $\Sigma$, so, actually, we are dealing with some sort of liquid. It is easy to check that
$$
C_0^2 = \frac{1}{2 \, r_0^2} \ ; \qquad \Delta p = \frac{2 \, \Sigma}{\sqrt 2 \, r_0} \ ;
$$
$$
e^{\nu} = \frac{1}{2} \left( \frac{r}{r_0} \right)^2 \ ; \qquad p_0 = \varepsilon_0 = \frac{1}{16 \, \pi \, G r_0^2} \ ;
$$
\begin{equation}
\label{mc}
m = m_0 = \frac{r_0}{4 \, G} \, .
\end{equation}
Note that the bare mass $M = \sqrt 2 \, m$, the relation is exactly the same as for the shell ``no-memory'' state, Eqn.~(\ref{sqrt2}), and $r_0 = 4 \, G m_0$, so, the size of our analog model is twice as that for a classical black hole of the same mass.

\medbreak

Now, how about the special point in our solution, $r=0$? It is not a trivial coordinate singularity, like in a three-dimensional spherically symmetric space, because
\begin{equation}
\label{hor}
ds^2 \, (r=0) = 0 \, .
\end{equation}
This looks like an event horizon. Indeed, the two-dimensional $(t-r)$-part of our metric describes a locally flat manifold. Since the static observers at $r = {\rm const}$ are, in fact, uniformly accelerated, this is a Rindler space-time with the event horizon at $r=0$. By definition, the surface of zero radius cannot be crossed, and this is just in this sense the generally global event horizon becomes local. The corresponding Rindler parameter which in more general case is called the ``surface gravity'', equals
\begin{equation}
\label{sgr1}
\varkappa = \frac{1}{2} \left\vert \frac{d\nu}{dr} \right\vert e^{\frac{\nu - \lambda}{2}} = \frac{C_0}{\sqrt 2} = \frac{1}{2 \, r_0} \, .
\end{equation}
Therefore, the Unruh temperature in our model is
\begin{equation}
\label{UT1}
T_U = \frac{1}{4 \, \pi \, r_0} = \frac{1}{16 \, \pi \, Gm} \, ,
\end{equation}
what is twice less than the Hawking temperature for the Schwarzschild black hole,
\begin{equation}
\label{HUT}
T_H = \frac{1}{8 \, \pi \, Gm} = 2 \, T_U \, .
\end{equation}
Note that the local observer measures the temperature
\begin{equation}
\label{loct}
T_{\rm loc} = \frac{1}{2 \, \sqrt 2 \, \pi \, r}
\end{equation}
which does not depend on the boundary value $r_0$, and is, in fact, universal.

\medbreak

To clarify the situation with the temperature, let us consider the general form of spherically symmetric line element with the Rindler two-dimensional part:
\begin{equation}
\label{R4}
ds^2 = a^2 \, \rho^2 \, dt^2 - d\rho^2 - R^2 (t,\rho) (d\theta^2 + \sin^2 \theta \, d\varphi^2) \, ,
\end{equation}
where $R(t,\rho)$ is the radius, and $a={\rm const}$ (the possible dependence $a(t)$ can always be absorbed by redefinition of the time coordinate). Then the Einstein equations read as follows
\begin{eqnarray}
\label{ee2}
G_0^0 &= &- \frac{2R''}{R} + \frac{\dot R^2}{a^2 \, \rho^2 \, R^2} + \frac{1-R'^2}{R^2} = 8 \, \pi \, G \, \varepsilon \nonumber \\
G_{01} &= &-2 \, \frac{\dot R'}{R} + 2 \, \frac{\dot R}{\rho \, R} = 8 \, \pi \, G \, T_{01} \nonumber \\
G_1^1 &= &\frac{2 \, \ddot R}{a^2 \, \rho^2 \, R} + \frac{\dot R^2}{a^2 \, \rho^2 \, R^2} - 2 \, \frac{R'}{\rho \, R} + \frac{1-R'^2}{R^2} = - 8 \, \pi \, G \, p_r \nonumber \\
G_2^2 &= &\frac{\ddot R}{a^2 \, \rho^2 \, R} - \frac{R''}{R} - \frac{R'}{\rho \, R} = - 8 \, \pi \, G \, p_t \, .
\end{eqnarray}
Then, demanding $T_{01} = 0$ because in thermal equilibrium there is no energy (heat) flow, we obtain
\begin{equation}
\label{rho}
R = \alpha (t) \, \rho \, .
\end{equation}
For static space-times $\alpha = {\rm const}$, and we recover the ``no-memory'' condition. We see that, in general, the radial and tangential pressures are not equal, $p_r \ne p_t$, and our liquid is anisotropic. But in the absence of external fields or some other influence (Coulomb force, angular momentum and so on) the local observer should see the isotropic surroundings. So, our assumption about perfect fluid seems correct.

\section{Thermodynamics}
\setcounter{equation}{0}

We saw that all the parts of our matter distribution are in thermal equilibrium. This is also reflected in the following remarkable feature. If one removes some outer layer, nothing would be changed inside. In this way the analog model resembles the (quasi-) classical black hole, where only the surface of the horizon is responsible for everything. Another feature is the existence of the intrinsic black hole frequency (resonance frequency) that follows from studying of quasi-normal modes. This forces us to consider the whole system as the set of quasi-particles, black hole phonons. The number of these phonons will be one of the thermodynamical extensive parameters, together with the entropy and the volume.

\medbreak

We are going to derive the local thermodynamical relations for our system. They should be distinguished from the global ones observed at infinity. The local observer deals with the bare mass $M$ ($=$ energy $E$) defined as the following integral over some volume $V$:
\begin{equation}
\label{defM}
E = M = \int T^{0\lambda} \, \xi_{\lambda} \, dV = \int T_0^0 \, \xi^0 \, dV = \int \varepsilon \, dV \, ,
\end{equation}
where $T_{\nu}^{\lambda}$ is the energy-momentum tensor, $\xi^{\mu}$ is the Killing vector normalized as $\xi^0 = 1$. In relativistic theory both the temperature $T$ and the entropy $S$ are temporal components of corresponding four-vectors. We will be using the local temperature $T_{\rm loc}$ and the invariant quantity $S = \xi_{\mu} \, S^{\mu}$, where $S^{\mu}$ is the entropy flow. With this in mind we can write the first law of thermodynamics as
\begin{equation}
\label{th1}
dM = \varepsilon \, dV = T_{\rm loc} \, dS - p \, dV + \mu \, dN \, ,
\end{equation}
$\mu$ is the chemical potential related to the number of black hole phonons (this is how the integer number enters our model), it ought to be included because in the model all the distributions are universal and the only parameter that changes is the boundary value of radius $r_0$, and this means automatical changing of all the extensive variables, $M$, $S$, $V$ and $N$. Dividing the above expression by the volume element $dV$ we get the first law in its local form
\begin{equation}
\label{th2}
\varepsilon (r) = T_{\rm loc} (r) \, s(r) - p(r) + \mu (r) \, n(r) \, ,
\end{equation}
where $s$ and $n$ are the entropy and particle densities, respectively. In our model $\varepsilon = p$, but how about $s$? The local observer cannot calculate it without knowing the corresponding microscopic structure, but he can ask his global counterpart who is educated enough (read proper books) and knows that the total entropy of the black hole is $S = \frac{1}{4 \, G} \, A_{\rm hor}$, what for the Schwarzschild black hole gives $(A_{\rm hor} = 4 \, \pi \, r_g^2)$ $S = \frac{\pi}{G} \, r_g^2 = \frac{\pi \, r_0^2}{4 \, G}$. Having this information, our local observer can deduce that
\begin{equation}
\label{ed}
s(r) = \frac{1}{8 \, \sqrt 2 \, Gr}
\end{equation}
and
\begin{equation}
\label{Ts}
T_{\rm loc} (r) \, s(r) = \frac{1}{32 \, \pi \, Gr^2} \, .
\end{equation}
Remembering now that $\varepsilon = \frac{1}{16 \, \pi \, Gr^2}$, we obtain
$$
T_{\rm loc} (r) \, s(r) = \frac{1}{2} \, \varepsilon \, , \quad \mu (r) \, n(r) = \frac{3}{2} \, \varepsilon \, .
$$

To construct a thermodynamical potential of energy $E = E(S,V,N)$, we make use of its additivity property,
\begin{eqnarray}
\label{add}
&E = N\varphi (x,y) \, , &x = \frac{S}{N} \, , \ y = \frac{N}{V} \, , \nonumber \\
&\varepsilon = y \varphi (x,y) \, , &p = - \frac{\partial \, E}{\partial \, V} = y^2 \, \frac{\partial \, \varphi}{\partial \, y} \, .
\end{eqnarray}
Since in our model $\varepsilon = p$, we obtain
\begin{equation}
\label{epsp}
\varphi = n \, \alpha (x) = y \, \alpha (x) \qquad p = n^2 \alpha (x) = y^2 \alpha (x) \, .
\end{equation}
Now, $T = \frac{\partial \, E}{\partial \, S} = \frac{\partial \, \varphi}{\partial \, x} = y \, \frac{\partial \, \alpha}{\partial \, x}$, $s = xy$, so $Ts = y^2 \, x \, \frac{\partial \, \alpha}{\partial \, x}$. Then, from the relation $Ts = \frac{1}{2} \, \varepsilon$ we get
\begin{equation}
\label{alpha}
x \, \frac{\partial \, \alpha}{\partial \, x} = \frac{1}{2} \, \alpha \, , \qquad \alpha = \alpha_0 \, \sqrt x \, ,
\end{equation}
where $\alpha_0$ is some universal constant. Eventually, we have
\begin{eqnarray}
&\varepsilon = \alpha_0 \, s^{1/2} \, n^{3/2} \, , &p = \alpha_0 \, s^{1/2} \, n^{3/2} = \varepsilon \, , \nonumber \\
&T = \frac{\alpha_0}{2} \, \frac{n^{3/2}}{s^{1/2}} \, , &\mu = \frac{3}{2} \, \alpha_0 \, s^{1/2} \, n^{1/2} \, . \nonumber
\end{eqnarray}
\begin{equation}
\label{thp}
E = \alpha_0 \, \frac{\sqrt{SN^3}}{V} \, .
\end{equation}

In what follows we will need the expression for the free energy $F$:
\begin{eqnarray}
\label{fe}
F &= &\int f \, dV \nonumber \\
f &= &\varepsilon - T_{\rm loc} \, s = \frac{1}{2} \, \varepsilon \nonumber \\
F &= &\frac{1}{2} \, M \, .
\end{eqnarray}
It is known that the thermal equilibrium conditions for the systems in static gravitational field are (see, i.e., \cite{LL})
\begin{eqnarray}
\label{tmu}
T \, \sqrt{g_{00}} &= &{\rm const} \, , \nonumber \\
\mu \, \sqrt{g_{00}} &= &{\rm const} \, .
\end{eqnarray}
The constants on the right-hand sides are universal for our model -- they do not depend on the boundary value $r_0$. Therefore, their ratio is also a universal constant. Thus, we have
\begin{equation}
\label{gamma}
\frac{\mu}{T} = 3 \, \frac{s}{n} = 3 \, \frac{S}{N} = 3  \, \gamma_0 \, .
\end{equation}
Hence, the entropy is naturally quantized:
\begin{equation}
\label{entrsp}
S = \gamma_0 \, N \, , \qquad N = 1,2 \ldots
\end{equation}
The constant $\gamma_0$ is, therefore, universal. It can be easily related to $\alpha_0$ that appeared in thermodynamical relations, Eqn.~(\ref{thp}). Indeed,
\begin{equation}
\label{calog}
\varepsilon = \alpha_0 \left( \frac{n}{s} \right)^{3/2} \, s^2 = \frac{\alpha_0}{\gamma_0^{3/2}} \, s^2 \, .
\end{equation}
Since $\varepsilon = \frac{1}{16 \, \pi \, Gr^2}$, and $s = \frac{1}{8 \, \sqrt 2 \, Gr}$, then
\begin{equation}
\label{calca}
\alpha_0 = \frac{8 \, G}{\pi} \, \gamma_0^{3/2} \, .
\end{equation}

\section{Solving the mystery of $\log 3$}
\setcounter{equation}{0}

In order to calculate the spacing coefficient $\gamma_0$ we have to make some assumption about the microscopic structure of our model. We assume that the interior matter distribution consists of $N$ black hole phonons with the equidistant spectrum of excitations
\begin{equation}
\label{phon}
\varepsilon_n = \omega \, n \, , \qquad n = 1,2 \ldots
\end{equation}
In this case the partition function for the whole system is the product of that ones for each phonons, and
\begin{eqnarray}
\label{pf1}
Z_{\rm tot} &= &(Z_1)^N \nonumber \\
Z_1 &= &\sum_n e^{-\frac{\varepsilon_n}{T}} = \sum_n \left( e^{-\frac{\omega}{T}} \right)^n = \frac{e^{-\frac{\omega}{T}}}{1-e^{-\frac{\omega}{T}}} \, .
\end{eqnarray}
It is natural to suppose that $\omega$ is just the black hole resonance frequency, its existence follows from the properties of quasi-normal modes (as was already explained earlier). Of course, $\omega$ is a temporal component of a four-vector, but the temperature $T$ also does, so their ratio does not depend on the choice of the clocks by local static observers. We accept that the observers are using their proper time, so $T$ is just the Unruh temperature $T_U$ which is constant in the whole interior. The partition function is an invariant, and we can calculate it in another way, using thermodynamical relations. Indeed, we can consider some small volume element $dV$ and the corresponding partition function $Z_{\rm small}$. Then, using the well-known formula for the free energy $F = -T \log Z$, and writing it for the volume element
\begin{equation}
\label{pf2}
dF = f \, dV = - T_{\rm loc} \log Z_{\rm small} \, ,
\end{equation}
where, as before, we use the local intrinsic quantities in thermodynamical relations. From this we have
\begin{equation}
\label{pf3}
\int \frac{f}{T_{\rm loc}} \, dV = - \sum \log Z_{\rm small} = -\log Z_{\rm tot} \, .
\end{equation}
The left-hand side equals
\begin{eqnarray}
\label{lhs}
\int \frac{f}{T_{\rm loc}} \, dV &= &\int \frac{\varepsilon - T_{\rm loc} \, s}{T_{\rm loc}} \, dV = \frac{1}{2} \int \frac{\varepsilon}{T_{\rm loc}} \, dV \nonumber \\
&= &2 \, \sqrt 2 \, \pi \int_0^{r_0} \frac{\varepsilon}{T_{\rm loc}} \, r^2 \, dr = \frac{\pi \, r_0^2}{4 \, G} = \frac{\pi \, r_g^2}{G} = S \, .
\end{eqnarray}
Here $r_g$ is the Schwarzschild radius, and $S$ is the total black hole entropy. Eventually, we obtain the important relation
\begin{equation}
\label{SZ}
e^{-S} = Z_{\rm tot} = (Z_1)^N \, ,
\end{equation}
from which it follows that
\begin{eqnarray}
\label{Sgamma}
\frac{e^{-\frac{\omega}{T}}}{1-e^{-\frac{\omega}{T}}} &= &e^{- \frac{S}{N}} = e^{-\gamma_0} \, , \nonumber \\
e^{\gamma_0} &= &e^{\frac{\omega}{T}} - 1 \, . 
\end{eqnarray}

To go further, let us consider the irreversible process of converting the mass (energy) of the system into radiation from a thermodynamical point of view. In our model such a process takes place just at the boundary $r=r_0$, and the thin shell with zero surface energy density and surface tension $\Sigma$ serves as a converter supplying the radiation with extra energy and extra entropy, this resembles the ``brick wall'' model \cite{tH}$^4$\footnotetext[4]{The nature of this radiation is purely quantum because our system is not radiating classically. The jump in the Unruh temperature of the inner and outer near-boundary static observers is compensated exactly by the gravitational influence of the surface tension.}. One can imagine that the near-boundary layer of thickness $\Delta \, r_0$ is converting into radiation, thus decreasing the boundary of the inner region to $(r_0 - \Delta \, r_0)$. Its energy equals $\Delta \, M = \varepsilon \, \Delta \, V$. To this we should add the energy released from the work done by the surface tension due to its shift, which is equal exactly to $\sum d (4 \, \pi \, r_0^2) = p \, d \, \Delta \, V = \varepsilon \, \Delta \, V = \Delta \, M$. Therefore, both the energy and the temperature in the converter becomes two times higher than that for any inner layer of the same thickness. And this double energy is gained by radiating quanta. Clearly, they have double frequency and exhibit double temperature, so
\begin{equation}
\label{log3}
\frac{R e \, w}{T_H} = \frac{\omega}{T_U} = \log 3 \, ,
\end{equation}
as follows from the spectrum of quasi-normal modes for the Schwarzschild black holes. Substituting this into Eqn.~(\ref{Sgamma}) and remembering that
\begin{equation}
\label{3-2}
3-1=2
\end{equation}
we obtain
\begin{equation}
\label{log2}
\gamma_0 = \log 2 \, .
\end{equation}
Since the radiated energy is thermalized, the interpretation of $dm$ as equal to $R e \, w$ is an improper procedure. This resolves the ``$\log 3$-paradox''.

\section{Acknowledgments}

The author is greatly indebted to the Institut des Hautes \'Etudes Scientifiques for kind hospitality extended to him. He would like to thank Thibault Damour, Pierre Vanhove, Maxim Kontsevich, Alexey Smirnov, Valentin Zagrebnov and Galina Eroshenko, Carlo Rovelli and members of his group for helpful discussions.

I am much thankful to C\'ecile Gourgues for careful typing the manuscript. Special thanks to my wife Anastasia Kupriyanova and to D.O., D.O. and N.O. Ivanovs for the permanent and strong moral support.

This work was supported by the grant No.~10-02-00635-a from the Russian Foundation of Fundamental Investigations (RFFI).

\end{document}